\documentclass[aps,prl,reprint,twocolumn,groupaddress,showpacs,amssymb]{revtex4-2}
\usepackage[english]{babel}
\usepackage{graphics}
\usepackage{graphicx}
\usepackage{epsfig}
\usepackage{amssymb}
\usepackage{dcolumn}
\usepackage{bm}
\usepackage{color}
\usepackage{natbib}
\usepackage{lineno}
\usepackage{sidecap}
\usepackage{verbatim}  
\usepackage{vector}  
\usepackage{wrapfig}
\usepackage{enumerate}
\usepackage{sidecap}
\usepackage{amsmath}
\usepackage{hyperref}

\begin{document}

\title{Experimental Realization of Room Temperature Topological Hall Effect in Kagome Antiferromagnet}

\author{Achintya Low}
\author{Susanta Ghosh}
\author{Sayan Routh}
\author{S. Thirupathaiah}
\email{setti@bose.res.in}

\affiliation{Department of Condensed Matter and Materials Physics, S . N. Bose National Centre for Basic Sciences, JD Block, Sector 3, Salt Lake,  Kolkata-700106}

\begin{abstract}
 Magnetic topological semimetals are the manifestations of interplay between electronic and magnetic phases of the matter, leading to peculiar characteristics such as the anomalous Hall effect (AHE) and topological Hall effect (THE). Mn$_3$Sn is a time-reversal symmetry broken (TRS) magnetic Weyl semimetal showing topological characteristics within the Kagome lattice network. In this study, for the first time, we uncover large and pure topological Hall effect in Mn$_3$Sn at the room temperature, which is gradually suppressed by Fe doping at the Mn site of Mn$_{3-x}$Fe$_x$Sn. We further identify that the topological properties of these systems are highly anisotropic. These findings promise the realization of potential topotronic applications at room temperature.
\end{abstract}

\maketitle

Hall effect is one of the pioneering discoveries in condensed matter physics~\cite{EH1991}, gained a lot of attentiveness both from the fundamental and technological applications~\cite{Popovic2004, Ramsden2006}. Specifically, in the fundamental science, the Hall effect branched into various exotic phenomena such as the spin Hall effect (SHE)~\cite{Dyakonov1971}, the quantum Hall effect (QHE)~\cite{Laughlin1981, Haldane1988, Smejkal2020}, the anomalous Hall effect (AHE)~\cite{Karplus1954}, and the recently emerging topological Hall effect in helimagnets. Particularly, the topological Hall effect is a direct manifestation of real-space Berry phase acquired by the conduction electrons moving around the topologically stable knots, also called the skyrmions, in chiral spin structures~\cite{Bruno2004,Lee2009, Neubauer2009}. Topological spin structures in helimagnets are driven by the geometrical frustration in triangular lattice or Dzyaloshinskii-Moriya interaction (DMI) in the absence of centrosymmetry of the crystal~\cite{Moriya1960,Shirane1983, Tomiyoshi1982a, Sticht1989, Nakatsuji2015, Park2018}. There exists several geometrically frustrated Kagome systems such as Co$_3$Sn$_2$S$_2$~\cite{Morali2019, Okamura2020}, Fe$_3$Sn$_2$~\cite{Baidya2020}, EuCd$_2$As$_2$~\cite{Sanjeewa2020} and  non-centrosymmetric crystals such as the cubic B20 chiral magnets MnSi~\cite{Muehlbauer2009}, FeGe~\cite{Yu2010}, MnGe~\cite{Tanigaki2015}, Cu$_2$OSeO$_3$~\cite{Janson2014}, and Mn$_2$RhSn~\cite{Rana2016} showing topological Hall effect. Importantly, most of these magnetic topological systems show THE besides the magnetization driven anomalous Hall effect hindering the potential applications of THE in the topological electronics~\cite{Shao2019, Luo2021, Gilbert2021}. Thus, finding materials merely with topological Hall effect is desirable for the technological applications.

\begin{figure*}
\centering
\includegraphics[width=1\linewidth]{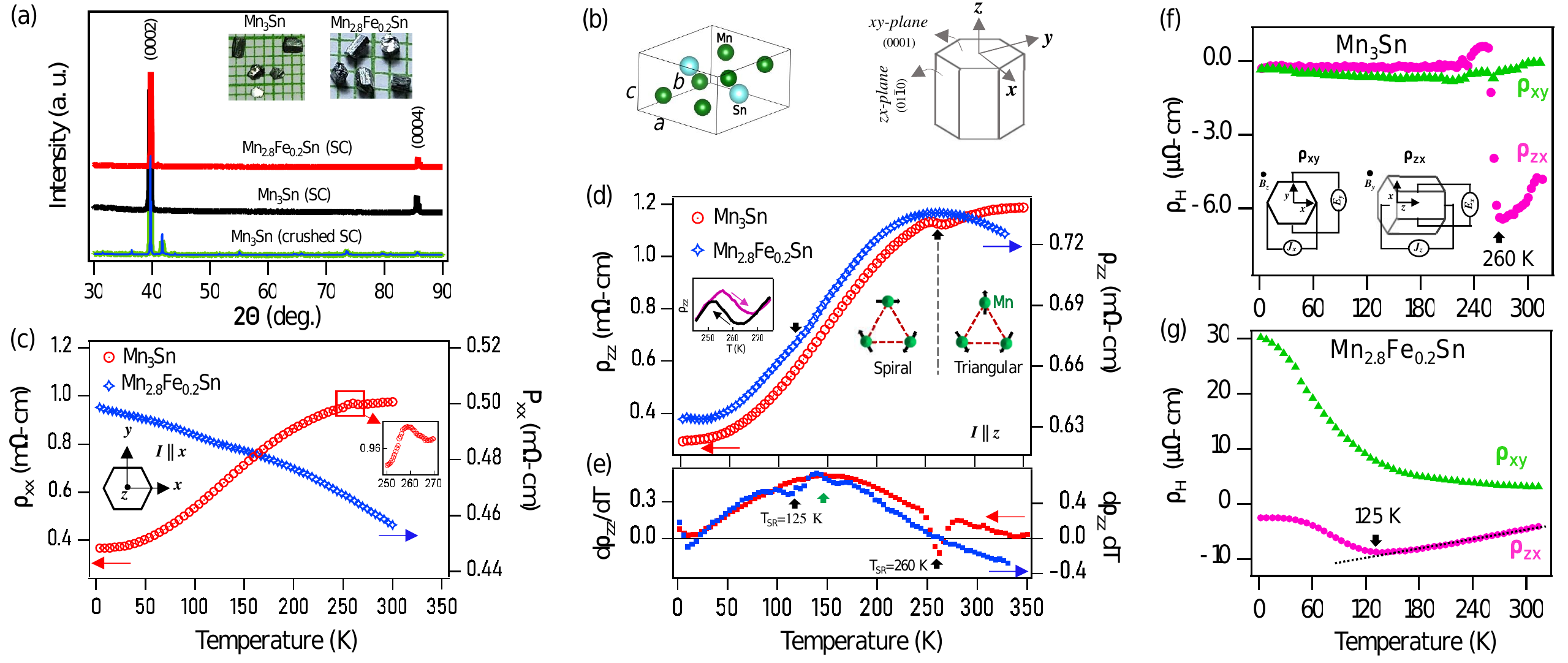}
\caption{(a) Powder XRD data from crushed single crystals of Mn$_3$Sn, single crystal of Mn$_3$Sn, and Mn$_{2.8}$Fe$_{0.2}$Sn.  Inset in (a) shows photographic image of  Mn$_3$Sn and Mn$_{2.8}$Fe$_{0.2}$Sn single crystals. Left-panel in (b) shows primitive unit cell of the hexagonal crystal structure of Mn$_3$Sn and the right-panel in (b) defines the crystal planes, $zx$-plane (01$\bar{1}$0) and $xy$-plane (0001) on the hexagonal unit cell. (c) Resistivity measured along the $a$-axis, $\rho_{xx}$,  as a function of temperature is plotted for both Mn$_3$Sn and Mn$_{2.8}$Fe$_{0.2}$Sn. (d) Resistivity measured along the $c$-axis, $\rho_{zz}$, as a function of temperature is plotted for both Mn$_3$Sn and Mn$_{2.8}$Fe$_{0.2}$Sn. Right-side inset in (d) shows schematic representation of the spin-reorientation above and below the transition temperature of 260 K for Mn$_3$Sn and 125 K for Mn$_{2.8}$Fe$_{0.2}$Sn. Left-side inset in (d) shows thermal hysteresis of the resistivity, $\rho_{zz}$ (T), between heating and cooling cycles of data collection on Mn$_3$Sn. (e) First derivative of $\rho_{zz}$ with respect to the temperature. (f) $xy$-plane ($\rho_{xy}$) and $zx$-plane ($\rho_{zx}$) Hall resistivity measured as a function of temperature from Mn$_3$Sn with a magnetic field of 1T applied parallel and perpendicular to the $c$-axis. Schematics in (f) show the measuring geometries used for recording the data of $\rho_{xy}$ and $\rho_{zx}$. (g) Similar data of (f) but measured from Mn$_{2.8}$Fe$_{0.2}$Sn.}
\label{1}
\end{figure*}

In this letter,  we report on the topological Hall effect of Mn$_{3-x}$Fe$_x$Sn (x=0, 0.2, 0.25, and 0.35) single crystals. Our studies show  a large and pure room-temperature topological Hall effect with topological Hall resistivity $\rho^{T}_{xy}$ $\approx$ 2 $\mu\Omega$-cm in Mn$_3$Sn at a low critical field of 0.3 T. To our knowledge, no other system shows such a pure and large THE signal at room temperature in its bulk form except Gd$_2$PdSi$_3$ which shows THE, with negligible AHE signal, but only at very low temperatures ($<$ 25K)~\cite{Kurumaji2019}.  The value of $\rho^{T}_{xy}$ increases further to 2.3 $\mu\Omega$-cm with decreasing temperature down to 260 K, but abruptly disappears below 260 K due to magnetic reorientation from coplanar AFM to spin-spiral structure~\cite{Sung2018, Yan2019}. We uncover highly anisotropic topological properties of these systems at room temperature. That means, anomalous Hall effect (AHE) has been noticed in the $zx$-plane, while the topological Hall effect has been noticed in $xy$-plane. Moreover, the room-temperature THE decreases with increasing Fe doping and it is almost suppressed by the Fe doping of $x$=0.35. Additionally, low temperature topological properties emerge by Fe doping. In the below, we explore in detail our experimental findings.

Single crystals of Mn$_{3-x}$Fe$_x$Sn were prepared by the self-flux method as reported in~\cite{Sung2018, Low2022}. See supplementary information for additional details on the experimental procedure and sample characterization. Electrical resistivity measured along the $a$-axis ($\rho_{xx}$) plotted as a function of temperature in Fig.~\ref{1}(c) for both Mn$_3$Sn and Mn$_{2.8}$Fe$_{0.2}$Sn. From Fig.~\ref{1}(c) we can observe that the $\rho_{xx}$ resistivity of the parent compound shows a metallic nature,  and interestingly for the first time, we find a hump-like structure at around 260 K [shown in the inset of Fig.~\ref{1} (c)], whereas, in the case of Fe doped Mn$_{2.8}$Fe$_{0.2}$Sn, the resistivity decreases with increasing temperature which is a kind of bad-metallic behaviour~\cite{Xu2020}. Similarly, electrical resistivity measured along the $c$-axis ($\rho_{zz}$) plotted as a function of temperature in Fig.~\ref{1}(d) for both Mn$_3$Sn and Mn$_{2.8}$Fe$_{0.2}$Sn. As can be seen from the $\rho_{zz}$  resistivity data, both parent and Fe doped systems show metallic behavior at low temperatures except for a significant increase in the impurity resistivity from 0.3 m$\Omega$-cm to 0.63 m$\Omega$-cm with Fe doping. Further, we observe a $kink$ at around 260 K in Mn$_3$Sn which has been ascribed earlier to the spin structure reorientation of Mn atoms from a high-temperature noncollinear inverse triangular structure to a low-temperature noncoplanar spiral structure~\cite{Sung2018, Yan2019}. But with Fe doping, we find that the spin reorientation transition temperature decreased to 125 K at which the $kink$ has been observed from the $\rho_{zz}$ resistivity of  Mn$_{2.8}$Fe$_{0.2}$Sn as shown in Fig.~\ref{1} (d). Inset in Fig.~\ref{1} (d) shows thermal hysteresis in the $\rho_{zz}$ resistivity of Mn$_3$Sn taken around 260 K between heating and cooling cycles. Thermal hysteresis in the $\rho_{zz}$ resistivity is consistent with the previous report on Mn$_3$Sn except that it was found at 270 K ~\cite{Yan2019} and 275 K~\cite{Sung2018}. Such a thermal hysteresis is attributed to the first-order type magnetic transition from triangular to spiral structure in this system~\cite {Sung2018}. On the other hand, we do not observe such a thermal hysteresis in the $\rho_{zz}$ resistivity of Mn$_{2.8}$Fe$_{0.2}$Sn despite having the magnetic transition at around 125 K.

To pinpoint the spin-reorientation transition temperature (T$_{SR}$), we plotted $d\rho_{zz}/dT$ as a function of temperature  as shown in Fig.~\ref{1} (e). From the first derivative,  we can reaffirm  T$_{SR}$=260 K for Mn$_3$Sn and 125 K for Mn$_{2.8}$Fe$_{0.2}$Sn. In addition, we also observe a decrease in $d\rho_{zz}/dT$ with increasing $T$ for both systems below 10 K  possibly due to weak local potentials at low temperatures~\cite{Barone2013} and above 143 K due to an electronic phase transition. Eventually, $d\rho_{zz}/dT$ becomes zero at 265 K and beyond this temperature, it is negative for Mn$_{2.8}$Fe$_{0.2}$Sn. That means,  Mn$_{2.8}$Fe$_{0.2}$Sn shows a metal-insulator (MI) transition at around 265 K. On the other hand, in Mn$_3$Sn, the MI transition seems to be happening at much-elevated temperatures as $d\rho_{zz}/dT$ approaches zero at around 345 K~\cite{Tomiyoshi1987}. To emphasize here, Mn$_3$Sn shows nearly isotropic resistivity between $a$ and $c$ axes, while Mn$_{2.8}$Fe$_{0.2}$Sn shows a large resistivity anisotropy.

\begin{figure*}[ht]
\centering
\includegraphics[width=1\textwidth]{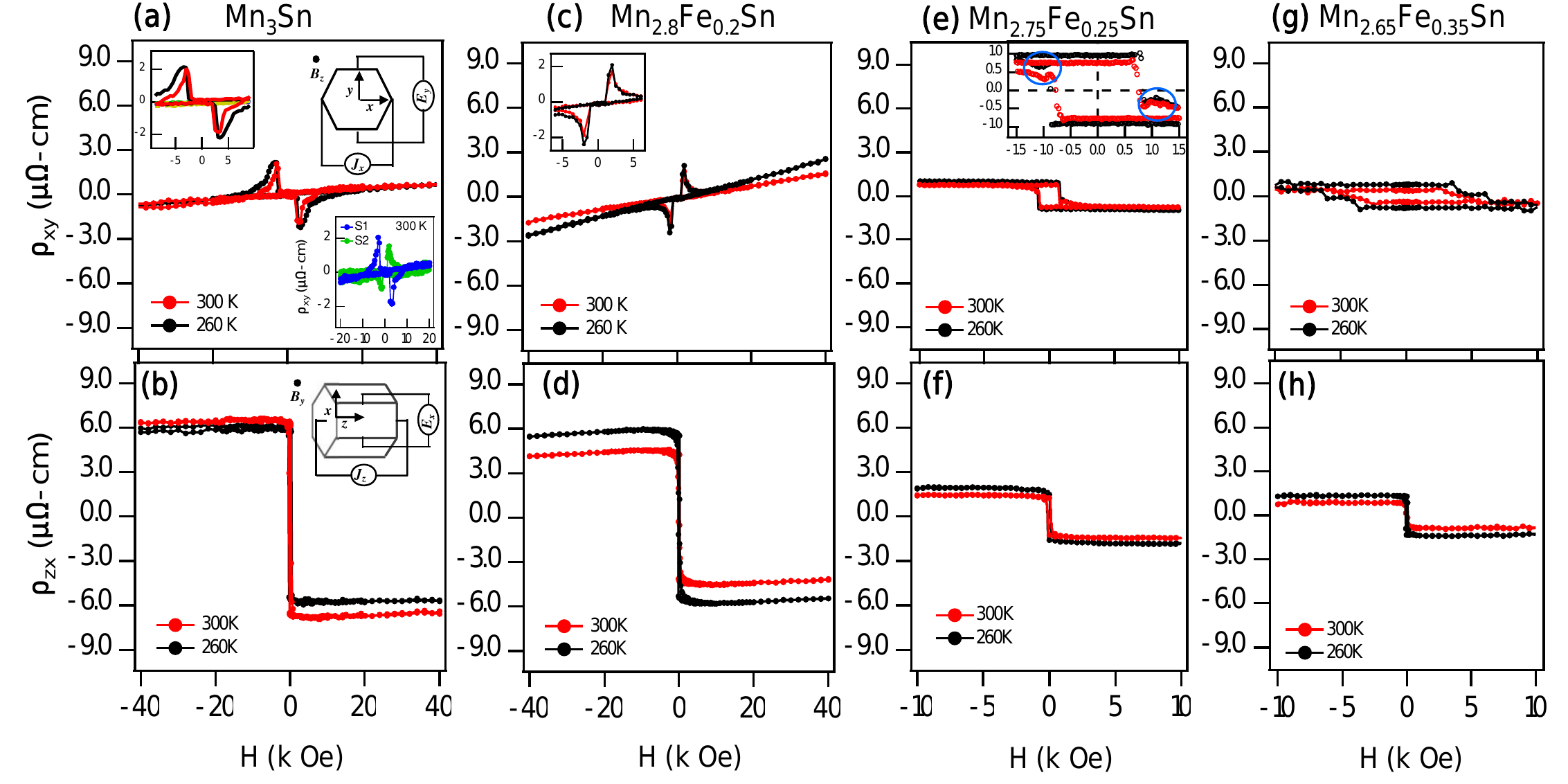}
\caption{(a) $xy$-plane ($\rho_{xy}$) Hall resistivity plotted as a function of the magnetic field from Mn$_3$Sn. The measuring geometry is shown in the top-right inset of (a) and zoomed-in data at low magnetic fields are shown in the top-left inset of (a). Bottom-right inset in (a) shows the Hall resistance (raw data) measured on two different samples (S1 and S2) of Mn$_3$Sn at 300K.   (b) $zx$-plane ($\rho_{zx}$) Hall resistivity plotted as a function of the magnetic field from Mn$_{3}$Sn. The measuring geometry is shown in the top-right inset of (b).  (c) and (d), (e) and (f), and (g) and (h) are similar Hall resistivity data of (a) and (b) but measured from Mn$_{2.8}$Fe$_{0.2}$Sn, Mn$_{2.75}$Fe$_{0.25}$Sn, and Mn$_{2.65}$Fe$_{0.35}$Sn, respectively.}
\label{3}
\end{figure*}

Fig.~\ref{1}(f) depicts Hall resistivity in the $xy$-plane ($\rho_{xy}$) and in the $zx$-plane ($\rho_{zx}$) plotted as a function of temperature from Mn$_3$Sn measured with a magnetic field of 1 T applied parallel and perpendicular to the $c$-axis. From Fig.~\ref{1}(f), we notice an increase in  $\rho_{zx}$ from 320 K down to 260 K, which then instantaneously become nearly zero below 260 K. This sudden decrease in $\rho_{zx}$ at 260 K coincides with the spin-reorientation transition temperature of the studied sample.  The observation of drastic changes in the $\rho_{zx}$ Hall resistivity at 260 K is in good agreement with previous studies on a similar system except for a slightly higher transition temperature of 270 K~\cite{Yan2019} and 275 K~\cite{Sung2018}. On the other hand, we observe no significant change in $\rho_{xy}$ from 320 K down to 2 K, which is nearly zero all the time. Next, Fig.~\ref{1}(g) depicts $\rho_{xy}$ and $\rho_{zx}$ Hall resistivity plotted as a function of temperature for Mn$_{2.8}$Fe$_{0.2}$Sn measured with magnetic field of 1 T applied parallel and perpendicular to the $c$-axis. From Fig.~\ref{1}(g), we notice an increase in  $\rho_{zx}$ from 320 K down to 125 K, which gradually decreases with temperature from 125 K down to 40 K, and then saturates below 40 K.  The decrease in $\rho_{zx}$ coincides with the spin-reorientation transition temperature of 125 K as observed from the magnetization measurements on the doped sample [see Fig.~S1(b) in the supplemental information]. On the other hand, $\rho_{xy}$ slowly increases from 320 K down to 125 K and exponentially increases from 125 K down to 40 K, and then tends to saturate below 40 K. Interestingly, we notice an extremely large $\rho_{xy}$ value of 31 $\mu \Omega-cm$ with 1 T field at 2 K which is negligible in Mn$_3$Sn.

Fig.~\ref{3} shows Hall resistivity, $\rho_{xy}(H)$ for $H\parallel c$ and $\rho_{zx}(H)$ for $H\perp c$, of  Mn$_{3-x}$Fe$_x$Sn measured at sample temperatures of 260 and 300 K. From Fig.~\ref{3}(a), we observe that the Hall resistivity ($\rho_{xy}$) of Mn$_3$Sn is as high as 2.3 $\mu\Omega$-cm at 260 K driven by a critical field of 0.3 T which is slightly reduced to 2 $\mu\Omega$-cm at 300 K. From $\rho_{zx}(H)$ of Mn$_3$Sn shown in Fig.~\ref{3}(b),  we observe anomalous Hall resistivity (AHR) as high as 6.5 $\mu\Omega$-cm at 300 K that in good agreement with previous studies on this system~\cite{Rout2019}. Next, from $\rho_{xy}(H)$ of Mn$_{2.8}$Fe$_{0.2}$Sn as shown in  Fig.~\ref{3}(c), we observe Hall resistivity as high as $~$2 $\mu\Omega$-cm at 260 K at a critical field of 0.19 T which then reduces to 1.6 $\mu\Omega$-cm at 300 K. Compared to Mn$_3$Sn, we notice significant decrease in AHR in Mn$_{2.8}$Fe$_{0.2}$Sn [see $\rho_{zx}(H)$ from Fig.~\ref{3}(d)], particularly at room temperature. Ideally, anomalous Hall effect is not expected in the in-plane Hall resistivity [$\rho_{xy}(H)$] of these systems~\cite{Nakatsuji2015}. But we find non-negligible anomalous Hall signal from the $\rho_{xy}(H)$ data of Mn$_{2.75}$Fe$_{0.25}$Sn [see Fig.~\ref{3}(e)],  possibly a component of $\rho_{zx}(H)$ [see Fig.~\ref{3}(f)] projected due to difficulties in making perfect Hall connections on the sub-millimeter sized crystals. However, from the zoomed-in $\rho_{xy}$(H) data of Mn$_{2.75}$Fe$_{0.25}$Sn [see inset in Fig.~\ref{3}(e)] we can clearly identify a cusp in Hall signal (marked by blue circles) at a critical field of 0.1 T. Similarly, Fig.~\ref{3}(g) shows $\rho_{xy}(H)$ of Mn$_{2.65}$Fe$_{0.35}$Sn in which again we find non-negligible anomalous Hall signal coming from $\rho_{zx}(H)$ [see Fig.~\ref{3}(h)]. Most importantly, we do not find any visible cusp in $\rho_{xy}$(H) data of Mn$_{2.65}$Fe$_{0.35}$Sn. Thus, as we go from Mn$_3$Sn to Mn$_{2.65}$Fe$_{0.35}$Sn we can clearly notice two things: i) The intensity of cusp in $\rho_{xy}(H)$ and the critical field at which the cusp appears decrease with increasing Fe doping and ii) Anomalous Hall signal in $\rho_{zx}(H)$ decreases with increasing Fe doping. Particularly, $\rho_{xy}(H)$ and $\rho_{zx}(H)$ of Mn$_{2.65}$Fe$_{0.35}$Sn are nearly suppressed at room temperature.

\begin{figure}[t]
\centering
\includegraphics[width=0.5\textwidth]{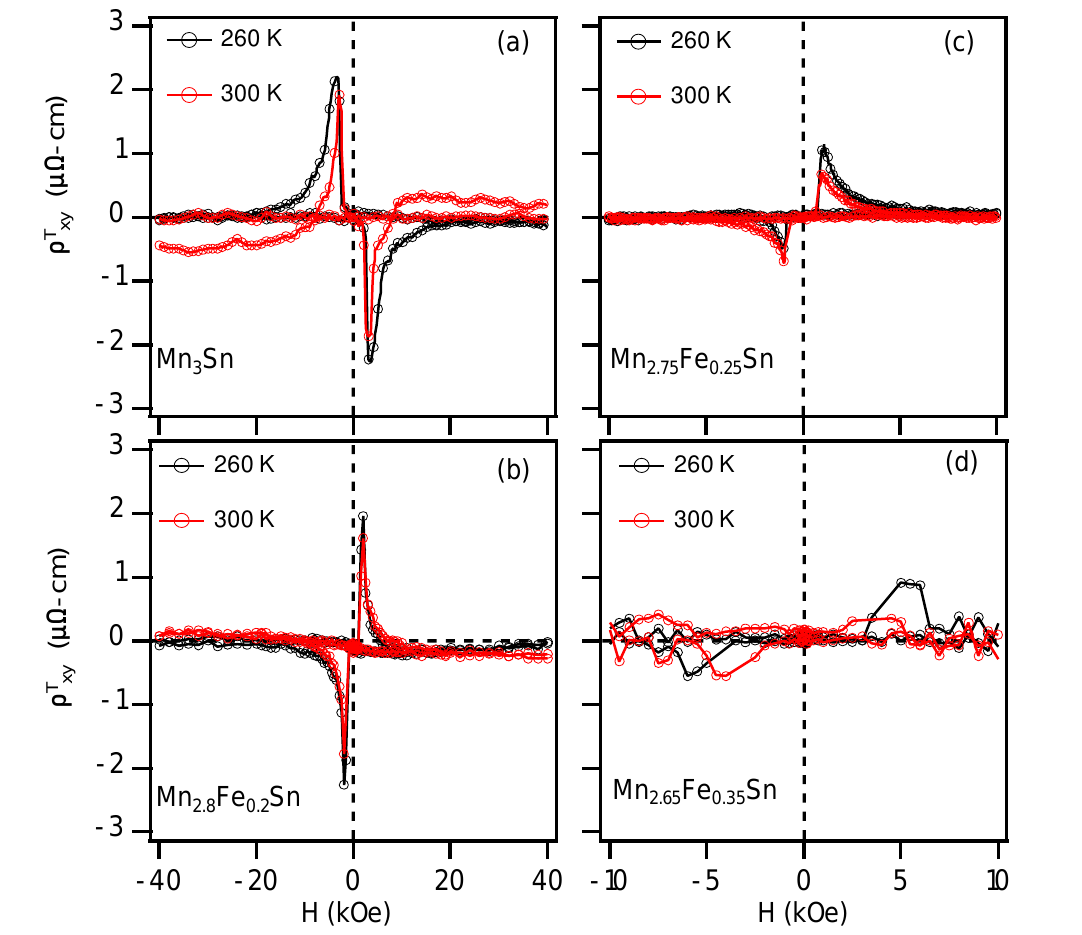}
\caption{$xy$-plane topological Hall resistivity ($\rho^T_{xy}$) plotted as a function of the field from Mn$_3$Sn (a), Mn$_{2.8}$Fe$_{0.2}$Sn (b), Mn$_{2.75}$Fe$_{0.25}$Sn (c), and Mn$_{2.65}$Fe$_{0.35}$Sn (d).}
\label{4}
\end{figure}

Next, coming to the important results of this study, Figs.~\ref{4}(a)-(d) show topological Hall resistivity (THR), $\rho^T_{xy}(H)$,  of Mn$_{3-x}$Fe$_{x}$Sn (x=0, 0.2, 0.25, and 0.35) extracted from the total Hall resistivity [$\rho_{xy}(H)$] shown in Figs.~\ref{3}(a), \ref{3}(c), \ref{3}(e), and \ref{3}(g), respectively. Note here that the total Hall resistivity shown in Fig.~\ref{3} has contributions from normal Hall resistivity ($\rho^N_{H}$)  which varies linearly with the field ($\rho^N_{H}=R_0 \mu_0 H$), anomalous Hall resistivity ($\rho^A_{H}$) which depends on the magnetization ($\rho^A_{H}=S_A \rho^2 M$), and topological Hall resistivity ($\rho^T_{H}$). All these contributions lead to a total Hall resistivity as per the relation $\rho_{H}=\rho^N_{H}+\rho^A_{H}+\rho^T_{H}=R_0 \mu_0 H+S_A\rho^2 M+\rho^T_{H}$. Here R$_0$ is a normal Hall coefficient and  S$_A$ is an anomalous Hall coefficient. To find the topological Hall resistivity one has to subtract the normal and anomalous Hall contributions from the total Hall resistivity,  following a method that has been explained thoroughly in several reports~\cite{Kanazawa2011, Nakatsuji2015, Kuroda2017, Rout2019}. However, since the anomalous Hall effect is negligible in $\rho_{xy}$ of these systems, we extracted $\rho^T_{xy}$ by simply subtracting the normal Hall contribution in the case of $x$=0 and $x$=0.2. For $x$=0.25 and $x$=0.35, we extracted  $\rho^T_{xy}$ by subtracting the anomalous Hall component originated from $\rho_{zx}$(H). Overall, Fig.~\ref{4} suggests that room temperature $\rho^T_{xy}$ is maximum in Mn$_3$Sn which then decreases with increase in doping and completely gets suppressed by $x$=0.35 of Fe doping. Moreover, the critical field at which the THE takes places also decreases with increase in doping.

\begin{figure}[t]
\centering
\includegraphics[width=0.5\textwidth]{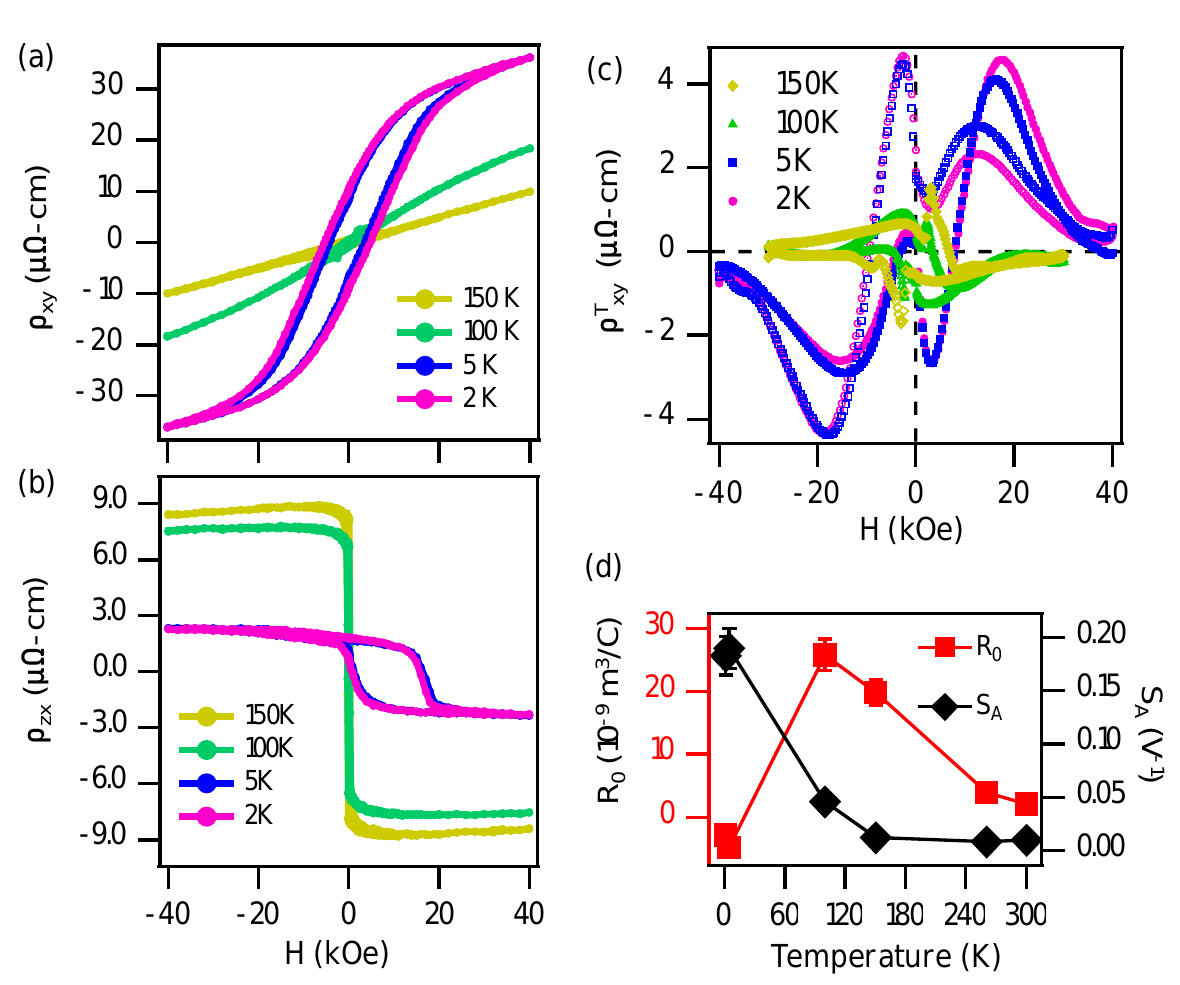}
\caption{Low temperature ($\leq$150 K) Hall resistivity of  Mn$_{2.8}$Fe$_{0.2}$Sn plotted as a function of field from the $xy$-plane (a) and $zx$-plane (b). (c) topological Hall resistivity ($\rho^T_{xy}$) plotted as a function of the field. (d) Normal Hall coefficient (R$_0$) and anomalous Hall coefficient (S$_A$) are plotted as a function of temperature (see the text for more details).}
\label{5}
\end{figure}

Fig.~\ref{5}(a) and ~\ref{5}(b) show $\rho_{xy}(H)$ and $\rho_{zx}(H)$ of Mn$_{2.8}$Fe$_{0.2}$Sn, respectively measured at low temperatures ($\leq$150 K). Interestingly, unlike in Mn$_3$Sn (see Ref.~\cite{Low2022} for low temperature data),  significant Hall resistivity was found in Mn$_{2.8}$Fe$_{0.2}$Sn as high as 36 $\mu\Omega$-cm at 2 K at an applied field of 4 T. We further observe hysteresis in $\rho_{xy}$(H) when measured at 2 and 5 K, which is consistent with our magnetization $M(H)$ data [see Fig.~S2(c) in the supplemental information]. This suggests that the anomalous Hall resistivity observed in Mn$_{2.8}$Fe$_{0.2}$Sn is mostly originated by the Fe-doping induced ferromagnetism~\cite{Low2022}. Further, unlike in Mn$_3$Sn,  from Fig.~\ref{5}(b),  we can see finite $\rho_{zx}$ in Mn$_{2.8}$Fe$_{0.2}$Sn even at 2 K which increases with temperature. We find $\rho_{zx}$$\approx$ 9.5 $\mu\Omega$-cm at 150 K. We also notice field-induced asymmetric hysteresis in $\rho_{zx}$ (H) at 2 and 5 K. Observation of asymmetric hysteresis in $\rho_{zx}(H)$  is consistent with the field-induced asymmetric $M(H)$ isotherm measured at 2 K [see Fig.~S2(d) in the supplemental information]. Fig.~\ref{5}(c) depicts $\rho^T_{xy}(H)$ extracted from Mn$_{2.8}$Fe$_{0.2}$Sn by following the above discussed technique. While Mn$_3$Sn shows no topological Hall resistivity in the $xy$-plane below 260 K, Mn$_{2.8}$Fe$_{0.2}$Sn shows topological Hall resistivity as high as 4.5 $\mu \Omega$-cm at 2 K, which gradually decreases with increasing temperature. But at 100 K, we notice that the sign of $\rho^T_{xy}$ switches from negative to positive for the positive magnetic fields and from positive to negative for the negative magnetic fields as shown in Figs.~\ref{5}(c). Fig.~\ref{5}(d) shows the normal Hall coefficient ($R_0$) and the anomalous Hall coefficient ($S_A$) plotted as a function of temperature, from which we  notice negative normal Hall coefficient at low temperatures,  suggesting for dominant electron carriers and at high temperatures $R_0$ becomes positive due to dominant hole carriers. The normal Hall coefficient sign switching is in-line with the sign change observed in the topological Hall resistivity [see Fig.~\ref{5}(c)]. Further, the anomalous Hall coefficient gradually decreases with increasing temperature from 0.18 V$^{-1}$ at 2 K to 0.01 V$^{-1}$ at 150 K and beyond 150 K $S_A$ is temperature independent.

Overall, the room-temperature pure and large topological Hall effect observed in  Mn$_3$Sn is an important discovery of this study as to date no other system shows $\rho^{T}_{xy}$ as high as 2 $\mu\Omega$-cm  at 300 K for such a low critical field of 0.3 T. Moreover, earlier studies on Mn$_3$Sn reported topological Hall resistivity, $\rho^{T}_{xy}$, in the range of 0.3 - 2.1 $\mu\Omega$-cm besides large anomalous Hall signal~\cite{Li2018, Li2019, Yan2019}. Also, a few systems other than Mn$_3$Sn show room-temperature THE but again with coexisting AHE. For instance, noncollinear ferromagnet LaMn$_2$Ge$_2$ shows topological Hall resistivity of $\approx$ 1$\mu\Omega$-cm, in addition to a comparable anomalous Hall resistivity of ~0.5 $\mu\Omega$-cm at 300 K~\cite{Gong2021}. Frustrated Kagome ferromagnet, Fe$_3$Sn$_2$, shows topological Hall resistivity of 2$\mu\Omega$-cm besides a large anomalous Hall resistivity of ~4.5 $\mu\Omega$-cm at 300 K~\cite{Wang2020}. Similarly, there exists many chiral and skyrmionic systems such as CrTe$_2$~\cite{Huang2021}, NiMnGa~\cite{Zhang2019},  Mn$_2$PtSn~\cite{Li2018a}, and YMn$_6$Sn$_6$~\cite{Ghimire2020} showing room-temperature topological Hall resistivity with a significant AHE signal. So far,  Gd$_2$PdSi$_3$ is the only system showing a large topological Hall resistivity of 2.5 $\mu\Omega$-cm,  with negligible AHE signal,  but at very low-temperatures ($<$ 25 K)~\cite{Kurumaji2019}. Interestingly, Mn$_{2.8}$Fe$_{0.2}$Sn also show pure room-temperature THE with $\rho^{T}_{xy}$$\approx$ 1.6 $\mu\Omega$-cm at a critical field of 0.2 T,  which then increases up to $\approx$ 4.5$ \mu\Omega$-cm with decreasing temperature down to 2 K  at a critical field of 1.7 T. Note here that the low temperature THE in Mn$_{2.8}$Fe$_{0.2}$Sn is coexisted with AHE. 

Topological Hall resistivity has been observed in many skyrmionic~\cite{Bruno2004, Lee2009, Neubauer2009, Kurumaji2019} and non-coplanar spin-structured systems~\cite{Shiomi2012,Roychowdhury2021,Gong2021,
Ghimire2020}. There exist various proposals to understand the stabilization of the skyrmion lattice in solids such as (i) DM interaction in noncentrosymmetric systems~\cite{Kanazawa2015, Park2018, Liu2017}, (ii)  uniaxial magnetocrystalline anisotropy in the centrosymmetric systems~\cite{Yu2014, Hou2018,Yu2012, Preissinger2021}, and (iii) in the systems with chiral domain walls~\cite{Gudnason2014, Cheng2019, Nagase2021, Yang2021}. Based on our experimental findings, we propose that the clean room-temperature $xy$-plane THE signal found in Mn$_{3-x}$Fe$_{x}$Sn for $x$=0, 0.2, 0.25 is originated either from the real-space Berry phase due to skyrmion lattice generated from the field-induced domain walls (DW)~\cite{Gudnason2014, Li2018, Li2019, Yan2019} or from the non-coplanar spin structure~\cite{Bruno2004, Lee2009, Neubauer2009, Kurumaji2019}. However, the low-temperature large THE found in Mn$_{2.8}$Fe$_{0.2}$Sn and in other Fe doped systems ($x$=0.25 and 0.35, reported separately in ~\cite{Low2022})  could be originated from the magnetocrystalline anisotropy induced by Fe doping~\cite{Yu2012,Hayami2016,Low2022}.

In summary, for the first time, we show a room-temperature pure in-plane topological Hall effect in Mn$_{3-x}$Fe$_{x}$Sn which decreases with increasing $Fe$ doping and is suppressed by a doping concentration of $x$=0.35. We show that the room-temperature topological properties are highly anisotropic in these systems. We also discovered that a small amount of Fe doping in Mn$_3$Sn causes dramatic changes in the low temperature topological properties. Additional experimental studies using Lorentz transmission electron microscopy (LTEM)  would be helpful to map the skyrmionic lattice in these systems. Moreover, our findings demand suitable theoretical models for understanding the high-temperature topological Hall effect in the Kagome lattice systems.

\section*{Acknowledgements}
Authors thank Science and Engineering Research Board (SERB), Department of Science and Technology (DST), India  for the financial support [Grant no. SRG/2020/000393].

\bibliography{Fe_Mn3Sn}

\nolinenumbers

\end{document}